\documentclass[conference]{IEEEtran}
\IEEEoverridecommandlockouts

\usepackage{cite}
\usepackage{amsmath,amssymb,amsfonts}
\usepackage{algorithmic}
\usepackage{graphicx}
\usepackage{textcomp}
\usepackage{xcolor}
\usepackage{microtype}

\usepackage{booktabs}
\usepackage{xcolor,colortbl}
\usepackage{siunitx}
\usepackage{hyperref}
\usepackage{multirow}
\usepackage{pifont}
\newcommand{\cmark}{\ding{51}}%
\newcommand{\xmark}{\ding{55}}%

\def\BibTeX{{\rm B\kern-.05em{\sc i\kern-.025em b}\kern-.08em
    T\kern-.1667em\lower.7ex\hbox{E}\kern-.125emX}}
\begin{document}

\title{Exploring Performance–Complexity Trade-Offs \\ in Sound Event Detection Models
\thanks{The LIT AI Lab is supported by the Federal State of Upper Austria. GW's work is supported by the European Research Council (ERC) under the European Union's Horizon 2020 research and innovation programme, grant agreement No 101019375 (Whither Music?).}
}


\author{
\IEEEauthorblockN{Tobias Morocutti$^2$, Florian Schmid$^1$, Jonathan Greif$^1$, Francesco Foscarin$^{1,2}$, Gerhard Widmer$^{1,2}$}
\IEEEauthorblockA{$^1$\textit{Institute of Computational Perception}, $^2$\textit{LIT Artificial Intelligence Lab}  \\
\textit{Johannes Kepler University}, Linz, Austria\\
\{tobias.morocutti, florian.schmid\}@jku.at}
}

\maketitle

\begin{abstract}

We target the problem of developing new low-complexity networks for the sound event detection task. Our goal is to meticulously analyze the performance-complexity trade-off, aiming to be competitive with the large state-of-the-art models, at a fraction of the computational requirements. We find that low-complexity convolutional models previously proposed for audio tagging can be effectively adapted for event detection (which requires frame-wise prediction) by adjusting convolutional strides, removing the global pooling, and, importantly, adding a sequence model before the (now frame-wise) classification heads. Systematic experiments reveal that the best choice for the sequence model type depends on which complexity metric is most important for the given application.
We also investigate the impact of enhanced training strategies such as knowledge distillation. In the end, we show that combined with an optimized training strategy, we can reach event detection performance comparable to state-of-the-art transformers while requiring only around $5\%$ of the parameters.
We release all our pre-trained models and the code for reproducing this work to support future research in low-complexity sound event detection\footnote{\href{https://github.com/theMoro/EfficientSED}{https://github.com/theMoro/EfficientSED}}.

\end{abstract}


\section{Introduction}

In recent years, the pursuit of better performance has led to ever larger deep learning models. 
The field of sound event detection (SED), i.e., the 
identification and exact temporal localization of specific sound events in audio files, has followed this trend, with state-of-the-art models reaching ~90M parameters~\cite{li24atst-frame,Shao24atst-sed,schmid2024multi,schmid2024EffectivePretrainingTransformerSED,Cai25MAT-SED}.
However, larger models incur significant inference costs, which can be prohibitive for applications with limited computational resources, such as embedded systems.


Prior research on low-complexity audio classification has leveraged efficient convolutional neural network (CNN) architectures adapted from the vision domain~\cite{Kong20PANNs,Gong21PSLA,schmid2023MobileNet} or introduced custom-designed efficient CNNs~\cite{Verbitskiy21ERANN,schmid2024dynamicMobileNet,kim2021broadcasted,cai2024tf,schmid2023distilling}. However, most of these models are optimized for clip-wise predictions, i.e., predicting a class label for an entire given audio clip. In contrast, the goal in SED is to identify specific acoustic events in a recording, including begin and end times, which implies a need for frame-wise predictions~\cite{Turpault19sed}. 

Unlike clip-wise tasks, where local features from the CNN are typically aggregated using global pooling, SED commonly employs a sequence model on top of the CNN to capture long-term temporal dependencies. Efficient SED architectures frequently utilize depthwise separable convolutions to reduce complexity~\cite{chan2023sed, drossos2020sed, Yang22lcsed}, yet the optimal choice of sequence model remains an open question. While recurrent neural networks (RNNs) are a widely adopted default—forming the basis of the popular convolutional recurrent neural network (CRNN) architecture~\cite{cakir17crnn, Xu18crnn, Wang19crnn, Li20crnn}—recent research has also shown promise in lightweight attention-based models~\cite{chan2023sed, Yang22lcsed}.

The present paper addresses this gap by systematically comparing various sequence models built upon a fixed CNN backbone, aiming to identify sequence models that offer the best trade-off between performance and complexity. We choose MobileNetV3~\cite{howard2019MobileNetV3} as the CNN backbone, as it is one of the most widely used depthwise separable CNNs and has demonstrated strong performance in large-scale audio tagging~\cite{schmid2023MobileNet}. By fixing the CNN architecture, we focus our analysis entirely on the temporal modeling stage.


To get the most meaningful results, we evaluate on the largest publicly-available SED dataset - \textit{AudioSet Strong}~\cite{audioset-strong} - and use the state-of-the-art training pipeline~\cite{schmid2024EffectivePretrainingTransformerSED}, which includes knowledge distillation~\cite{Hinton15kd} from large transformer models. Our experiments aim to answer the following key questions:
\begin{itemize}
    \item Which sequence model, when varying its size, yields the best performance-complexity trade-off? (Section~\ref{sec:exp_seq})
    \item Do the benefits of using a sequence model hold across multiple sizes of the CNN backbone? (Section~\ref{sec:model_scaling})
    \item How important are training techniques, such as knowledge distillation, for low-complexity SED? (Sections~\ref{sec:ablation} and~\ref{sec:pretraining_kd})
\end{itemize}

\section{Architectures}
\label{sec:architectures}

We assume the input of our task to be 10-second audio files, encoded as mel spectrograms with 128 frequency bins and 1000 time frames. 

All the architectures we propose consist of 3 main components: a frame-wise convolutional backbone $g$ (see Section~\ref{sec:conv_backbone}) that inputs the audio spectrogram $\mathbf{x}$, a sequence model $f$ (see Section~\ref{sec:sequence_model}) that processes the output $\mathbf{\hat{z}}$  of the backbone, and the task heads that produce the frame-wise sound event predictions $\mathbf{\hat{o}}_C$.

Formally, our model can be represented as:
\begin{displaymath}
\label{eq:final}
\begin{split}
 \mathbf{\hat{z}} & = g(\mathbf{x}) \\
 \{\mathbf{\hat{o}}_C\}_{t=1}^{250} & = \mathbf{W_h} f ( \mathbf{W_d} \{\mathbf{\hat{z}}\}_{t=1}^{250} + \mathbf{b_d}) + \mathbf{b_h} 
\end{split}
\end{displaymath}
\noindent
where $\mathbf{W_d}$ and $\mathbf{b_d}$ represent a linear projection aligning embeddings with the dimension of the sequence model $f$. If no sequence model is used, $f$ and the linear projection ($\mathbf{W_d}$, $\mathbf{b_d}$) are identity functions. The output of $f$ is then transformed by a position-wise linear layer, parameterized by $\mathbf{W_h}$ and $\mathbf{b_h}$, to generate frame-level predictions $\{\mathbf{\hat{o}}_C\}_{t=1}^{250}$, with dimension $C$ corresponding to the number of classes in the task. 

\subsection{Frame-wise Convolutional Backbone}
\label{sec:conv_backbone}

As the convolutional backbone, we use MobileNetV3~\cite{howard2019MobileNetV3}.
To enable frame-wise instead of clip-wise predictions, we remove the global pooling and the global classification heads. The outputs of this network are then three-dimensional embeddings $\mathbf{\hat{z}}$ of shape $\textit{channel} \times \textit{freq} \times \textit{time}$.
Compared to the original MobileNet, we increase some convolutional strides along the frequency axis from 1 to 2 while decreasing others along the time axis from 2 to 1. 
This changes the shape of $\mathbf{\hat{z}}$ from $\textit{channels} \times \textit{4} \times \textit{32}$ to $\textit{channels} \times \textit{1} \times \textit{250}$, matching the temporal resolution of the pseudo-labels used for knowledge distillation as described in Section \ref{sec:exp_settings}. Therefore, each of the 250 time frames in $\mathbf{\hat{z}}$ corresponds to a 40-millisecond segment of the original 10-second audio clip.
We refer to this adapted model as \textit{frame-wise MobileNet} (\textit{fmn}).

For the experiments that require it, we scale the frame-wise MobileNets following the method described in \cite{schmid2023MobileNet}, maintaining the number of layers while adjusting the model's width
(by multiplying the default number of channels of all convolutional layers) 
using a scaling factor $\alpha$. 
We consider models with $\alpha \in [0.4, 0.6, 1.0, 2.0, 3.0]$ which we name \textit{fmn04}, \textit{fmn06}, \textit{fmn10}, \textit{fmn20} and \textit{fmn30}, respectively.

\subsection{Sequence Models}\label{sec:sequence_model}

To investigate whether sequence models can improve the performance-complexity trade-off of low-complexity CNNs, we evaluate a range of well-established as well as recently proposed sequence models. Our experiments include transformer blocks (TF)~\cite{vaswani2017transformer}, multi-head self-attention (ATT)~\cite{vaswani2017transformer}, bidirectional gated recurrent units (BiGRU)~\cite{cho2014gru}, temporal convolutional networks (TCN)~\cite{oord2016wavenet}, Mamba blocks (MAMBA)~\cite{gu2023mamba, dao2024mamba2}, and a hybrid sequence model (HYBRID) that combines a minimal version of a GRU (minGRU)~\cite{feng2024minGRU} with a self-attention layer. 

For TF and ATT, we incorporate rotary positional embeddings~\cite{su2024roformer}, because these performed better than other positional encodings for small hidden dimensions in our preliminary studies. The TCN blocks consist of five convolutional layers with exponentially increasing dilation values, where the $n$-th layer has a dilation of $2^n$. TCN's model capacity is defined by the number of convolutional channels, analogous to the hidden dimension in recurrent models and the model dimension in attention-based and state-space models. For simplicity, we refer to the sequence model dimension as the hidden dimension throughout this work. For MAMBA, we use the more recent Mamba-2 blocks~\cite{dao2024mamba2} with a state dimension of 64. 

HYBRID consists of two parallel branches, following the design of Hymba~\cite{dong2024hymba}. 
However, inspired by \cite{minGRU2024}, we replace Mamba with a bidirectional minGRU~\cite{feng2024minGRU} and use it in parallel with self-attention.\footnote{
In HYBRID, we use bidirectional minGRUs instead of BiGRUs, as they require fewer parameters and can be efficiently parallelized using the Heinsen Parallel Scan Algorithm~\cite{heinsen2023parallelScan}. Since our focus is on offline inference, this algorithm can be leveraged for inference, significantly increasing throughput. 
In contrast to the unidirectional minGRUs implemented in \href{https://github.com/lucidrains/minGRU-pytorch}{https://github.com/lucidrains/minGRU-pytorch}, we employ bidirectional minGRUs for improved performance in our application.
} 
The outputs of these branches are averaged and linearly projected before serving as input to the subsequent sequence model block or the task head. 


\section{Experimental Setup}
\label{sec:exp_setup}
In this section, we outline our dataset, training configurations, performance metrics, and the methods used to assess the complexity of the proposed models.

\subsection{Dataset}
AudioSet Strong~\cite{audioset-strong} includes predefined training and evaluation sets containing 103,463 and 16,996 ten-second audio clips, respectively. Of these, we successfully downloaded 100,911 training clips and 16,935 evaluation clips. We train on all 447 available sound classes in the training set, while, in line with~\cite{li24atst-frame, schmid2024EffectivePretrainingTransformerSED}, we evaluate on the 407 classes that overlap between the training and evaluation sets.

\subsection{Training Configuration}
\label{sec:exp_settings}

We initialize the backbone CNNs with ImageNet~\cite{Deng09ImageNet} pre-trained weights and train them on AudioSet Weak~\cite{gemmeke2017AudioSet}, following the knowledge distillation routine from \cite{schmid2023MobileNet}. However, we find that extending the training to 300 epochs improves performance.

Following the AudioSet Weak pre-training, the full models (pre-trained backbone + sequence model) are trained on AudioSet Strong for 120 epochs using an Adam optimizer with no weight decay, a cosine learning rate schedule with 5,000 warmup steps, and a batch size of 256.

Our preliminary experiments showed that the best performance was achieved by using different learning rates for the pre-trained backbone and the randomly initialized sequence model. Specifically, for the backbone, we use a layer-wise learning rate decay of 0.9 and choose the learning rate with independent grid searches per model size over 1e-3, 3e-3, and 6e-3 without using any sequence model. We then select the sequence model learning rate independently for every type and size from a small, pre-defined grid ($\{\text{5e-4, 8e-4, 3e-3}\}$). 
Following \cite{schmid2024EffectivePretrainingTransformerSED}, we train our models using knowledge distillation on pseudo-labels generated by an ensemble of 15 transformers, in addition to the typical one-hot encoded labels. The distillation and hard label losses are weighted at 0.9 and 0.1, respectively. Furthermore, the frame-wise MobileNet is designed to align with the resolution (40 ms) of these pseudo-labels. Training our low-complexity models on the pseudo-labels enables a simple training setup without overfitting, using mixup~\cite{Zhang18mixup} as the only data augmentation.


To ensure a fair comparison among the sequence models, we optimize the median filter window size for postprocessing on \textit{fmn10} using no sequence model and apply the same setting across all experiments. Odd, centered window sizes yield significantly better performance than even ones, with a size of 9 frames (0.36 seconds) leading to the best results for \textit{fmn10}.  



\subsection{Performance and Complexity Metrics}
Following \cite{schmid2024EffectivePretrainingTransformerSED}, we assess performance using PSDS1~\cite{Ebbers2022psds}, a threshold-independent metric that quantifies the intersection between ground truth and detected events. By prioritizing low reaction time and precise event localization, PSDS1 aligns well with our primary research focus.

We use three measures of model complexity: the parameter count, the number of multiply-accumulate (MAC) operations\footnote{We use the model profiler contained in Microsoft's \href{https://www.microsoft.com/en-us/research/project/deepspeed/}{DeepSpeed framework} to measure MACs.}, and throughput, defined as the number of samples processed per second during inference. Throughput and MACs are computed using ten-second audio clips.
Throughput is measured on an Nvidia RTX 3090 GPU with a batch size of 64 and 16-bit floating-point precision, considering the PyTorch implementation of the models, which we publicly release.

\section{Results}
\label{sec:results}
\subsection{Performance-Complexity Trade-Off of Sequence Models}
\label{sec:exp_seq}

In this experiment, we investigate the performance-complexity trade-off of different sequence models built on top of \textit{fmn10}. We vary their size across four values of hidden dimension (128, 256, 512, and 1024) and compare their performance, total parameter count, throughput, and MACs.  

We explore stacking multiple sequence model blocks and observe that, across different sequence models, using two blocks results in more stable training and a notable performance improvement over a single block. However, adding a third block further increases the complexity without delivering a meaningful performance gain. Therefore, we consistently stack two blocks of the sequence model on top of the convolutional backbone.

\begin{figure}[h!]
    \centering
    \includegraphics[width=0.9\linewidth]{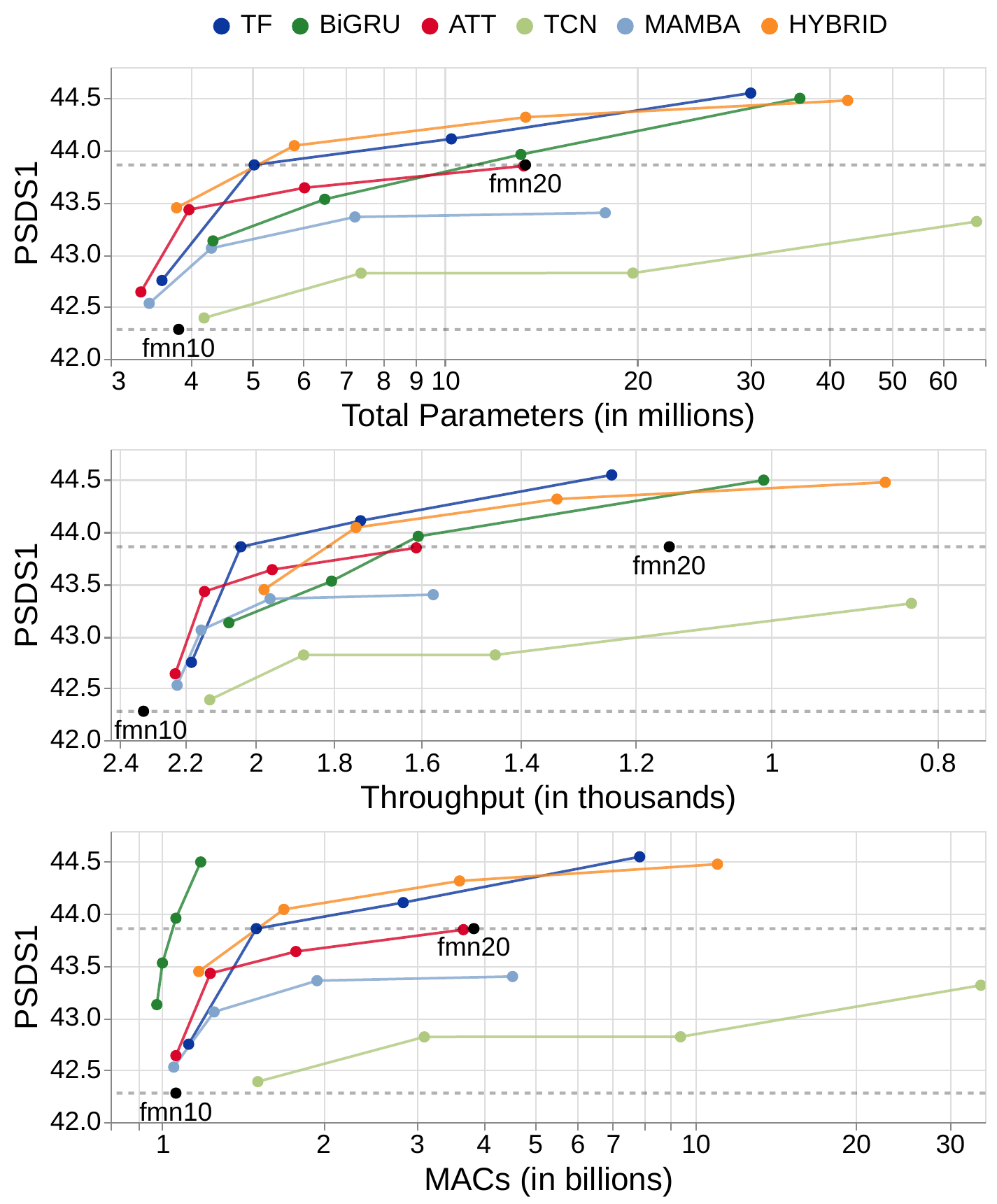}
    \caption{\textbf{Performance-Complexity trade-off} of \textit{fmn10} trained with various sequence models, scaled across four complexity levels. Colored points indicate increasing hidden dimensions (128, 256, 512, and 1024) from left to right. Evaluated sequence models include transformer blocks (TF), BiGRUs, self-attention layers (ATT), temporal convolutional networks (TCNs), Mamba blocks (MAMBA), and a hybrid sequence model (HYBRID) combining minGRU and self-attention. For reference, \textit{fmn10} and \textit{fmn20} without sequence models are shown as well.}
    \label{fig:plot_1}
    \vspace{-6pt}
\end{figure}

Figure \ref{fig:plot_1} shows that all sequence models improve the performance of \textit{fmn10}, even with the smallest hidden dimension. However, only models incorporating TF, BiGRU, ATT, and HYBRID achieve performance comparable to or surpassing \textit{fmn20} while maintaining lower parameter counts, lower MACs, and increased throughput, resulting in a more favorable performance-complexity trade-off. 
This shows that adding suitable sequence models on top of \textit{fmn10}
scales more efficiently than simply scaling up the convolutional backbone itself. In contrast, MAMBA and TCN fail to reach the performance level of \textit{fmn20} at any hidden dimension size. 


The choice of the model offering the best performance-complexity trade-off depends on the complexity metric of interest. The top choices for parameters, throughput, and MACs are HYBRID, TF, and BiGRU, respectively.
For example, \textit{fmn10} using TF and a hidden dimension of 256 (\textit{fmn10+TF:256}) matches the performance of \textit{fmn20} while achieving 1.76 times its throughput and requiring only 37.6$\%$ of its total parameters. Similarly, \textit{fmn10} with HYBRID at a hidden dimension of 128 (\textit{fmn10+HYBRID:128}) has fewer parameters than \textit{fmn10} without a sequence model\footnote{
If the down-projection, sequence model, and linear head together require fewer parameters or MACs than a single large linear layer mapping embeddings to target classes, adding a small sequence model can be more efficient than using no sequence model with respect to these metrics.
} yet improves the PSDS1 score from 42.28 to 43.45. However, \textit{fmn10+HYBRID:128} exhibits lower throughput compared to \textit{fmn10+TF:256}. 
In terms of MACs, BiGRU is the most effective\footnote{BiGRUs involve multiple operations, such as gating, point-wise products, and point-wise summations, which do not count as MACs.},
but its sequential processing limits parallel computation, resulting in lower throughput compared to TF and HYBRID. 
We conclude that TF is the most balanced choice with strong results on all three complexity measures. In particular, \textit{fmn10+TF:256} strikes an excellent trade-off between performance gain and minimal added complexity, making it the preferred model for further experiments.

\subsection{Scaling the Backbone and Sequence Model}\label{sec:model_scaling}

In the previous section, we analyzed the performance-complexity trade-off of sequence models using a fixed-size convolutional backbone (\textit{fmn10}). Here, we extend this analysis by jointly scaling both the convolutional backbone and the sequence model. As outlined in Section \ref{sec:architectures}, the frame-wise MobileNets are scaled by a factor $\alpha$, which we also apply to the sequence models' hidden dimension. Based on the findings in Section \ref{sec:exp_seq}, we set $\alpha=1.0$ to correspond with a hidden dimension of 256 and evaluate the three top-performing sequence models: TF, HYBRID, and BiGRU.

\begin{figure}[t]
    \centering
    \includegraphics[width=0.9\linewidth]{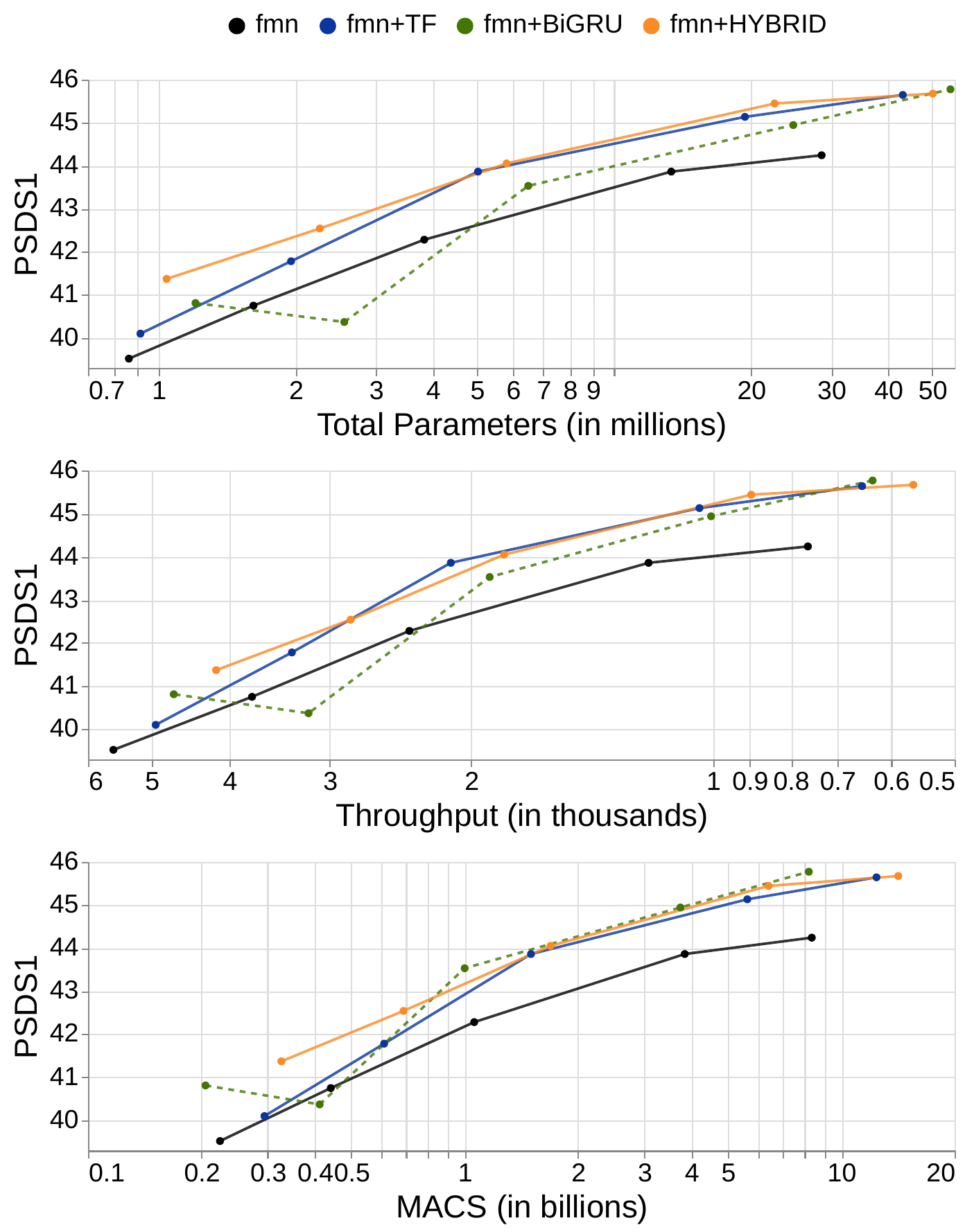}
    \caption{\textbf{Performance-Complexity trade-off} of the models scaled to different complexity levels. The points on the black line, from left to right, correspond to the models \textit{fmn04}, \textit{fmn06}, \textit{fmn10}, \textit{fmn20}, and \textit{fmn30}. We compare these to models incorporating transformer blocks (\textit{fmn+TF}), HYBRID (\textit{fmn+HYBRID}), or BiGRUs (\textit{fmns+BiGRU}) with sequence models scaled proportionally to the backbone. For clarity, the BiGRU line is dashed.}
    \label{fig:plot_2}
    \vspace{-6pt}
\end{figure}

Three key conclusions can be drawn from Figure~\ref{fig:plot_2}: 1) TF and HYBRID consistently outperform fmns without sequence models across all scaling factors; 
2) HYBRID performs best in the lower complexity range, with \textit{fmn04+HYBRID} and \textit{fmn06+HYBRID} surpassing their TF and BiGRU counterparts; and 3) TF and HYBRID exhibit greater stability than BiGRU, as the performance drop of BiGRU from $\alpha=0.4$ to $\alpha=0.6$ is a consistent trend across multiple independent runs. 

\subsection{Impact of KD and CNN Backbone Fine-Tuning}\label{sec:ablation}

\begin{table}[t]
\renewcommand{\arraystretch}{1.0}
\caption{Impact of KD and CNN Backbone Fine-Tuning. \label{tab:kd_tf_ablation}}
\centering
\begin{tabular}{lccc}
\toprule 
\textit{Model} & KD & Frozen CNN Backbone & PSDS1 \\
\midrule
\textit{fmn10} & \xmark  & \xmark & 38.11 \\
\textit{fmn10} & \cmark  & \xmark & 42.28 \\
\textit{fmn10+TF:256} & \xmark  & \xmark & 41.13 \\
\textit{fmn10+TF:256} & \cmark  & \xmark & \textbf{43.86} \\
\textit{fmn10+TF:256} & \cmark  & \cmark & 42.05 \\
\bottomrule
\end{tabular}
\vspace{-6pt} 
\end{table}

\renewcommand{\arraystretch}{1.1}

As outlined in Section \ref{sec:exp_settings}, 
we first pre-train the CNN backbone on AudioSet Weak and fine-tune the full model (backbone + sequence model) on AudioSet Strong using KD.
Table \ref{tab:kd_tf_ablation} presents the resulting performance effects of this setup.
Regardless of whether a sequence model (\textit{TF:256} on top of \textit{fmn10}) is used, KD significantly enhances performance. Similarly, fine-tuning the convolutional backbone for frame-wise SED predictions further improves results, though its impact is smaller than that of KD.


\subsection{Expanding Knowledge Distillation to AudioSet Weak}\label{sec:pretraining_kd}

This section explores techniques to further enhance the proposed low-complexity models, narrowing the gap to large state-of-the-art transformers. So far, we have followed the KD training routine from~\cite{schmid2024EffectivePretrainingTransformerSED}, which uses pre-computed transformer ensemble pseudo-labels on the AudioSet Strong training set. However, the distillation process is not limited to frame-wise annotated data (AudioSet Strong) and can be extended to other recordings. Following this idea, we leverage the top-performing transformer model on AudioSet Strong from~\cite{schmid2024EffectivePretrainingTransformerSED}, BEATs~\cite{Chen23BEATs}, to generate frame-level predictions for the AudioSet Weak training split ($\sim$2 million files). The distillation loss is then computed on batches containing 50\% AudioSet Weak and 50\% AudioSet Strong clips.

\begin{table}[t]
\renewcommand{\arraystretch}{1.1}
\caption{Improved Knowledge Distillation Variants.}
\label{tab:online_teacher}
\centering
\begin{tabular}{lccc}
\toprule
\textit{Model}  & Weak & Extended Training & PSDS1 \\
\midrule
\textit{fmn10+TF:256} & \xmark & \xmark & 43.86 \\  
\textit{fmn10+TF:256}  & \cmark & \xmark & 44.70 \\ 
\textit{fmn10+TF:256}  & \cmark & \cmark & \textbf{45.25} \\ 
\bottomrule
\end{tabular}
\vspace{-6pt} 
\end{table}
\renewcommand{\arraystretch}{1.0}

Table~\ref{tab:online_teacher} shows that incorporating AudioSet Weak for distillation boosts the performance of \textit{fmn10+TF:256} by nearly 2\%, increasing its PSDS1 score from 43.86 to 44.70. Given the large scale of AudioSet Weak, we extend training from 120 to 240 epochs, yielding an additional 1.2\% improvement and reaching a PSDS1 score of 45.25. These results highlight that an efficient model like \textit{fmn10+TF:256}, combined with an optimized training strategy, can achieve performance comparable to state-of-the-art transformers (PSDS1 scores between 45.4 and 46.5 in~\cite{schmid2024EffectivePretrainingTransformerSED}) while using only 5 million parameters—significantly fewer than the $\sim$90 million in transformer models\footnote{This enhanced distillation procedure is not applied to all experiments in this paper due to substantially higher training costs.}.

\section{Conclusion}



In this paper, we proposed and thoroughly evaluated efficient sound event detection (SED) architectures. We utilized modified MobileNetV3 models, adjusted to various complexity levels, as the convolutional backbone, and combined them with different sequence models. Our findings demonstrate that pairing appropriate sequence models with the backbone consistently outperforms fully convolutional versions across all complexity levels tested. 
The optimal sequence model depends on the prioritized complexity metric.
We further emphasized the critical role of knowledge distillation in achieving high performance and showed that, with an optimized distillation strategy, our models can match state-of-the-art transformer performance while using only around 5\% of their parameters. To aid future work, we release our pre-trained models along with our training and evaluation framework, inviting researchers to develop new sequence models that push the performance–complexity trade-off even further.

\vspace{-3pt}

\bibliographystyle{IEEEtran}
\bibliography{refs}

\end{document}